\documentclass[%
aps, %
prl, floatfix, 
twocolumn, 10pt, 
a4paper, %
nofootinbib, %
amsfonts, amsmath, amssymb]{revtex4-2}
\usepackage[sort&compress]{natbib}
\usepackage{newtxtext}
\usepackage[varg]{newtxmath}
\usepackage{ascmac}
\usepackage{mathrsfs}
\usepackage{bm}
\usepackage{color}
\usepackage{graphicx}
\usepackage{bm}
\usepackage{braket}
\usepackage[normalem]{ulem}
\usepackage{hhline}
\usepackage{hyperref}
\hypersetup{%
    unicode=false, %
    bookmarksnumbered=true, %
    bookmarkstype=toc, %
    breaklinks=true, %
    colorlinks=true, %
    pdfborder={0 0 1}, %
    linkcolor=blue, %
    citecolor=blue, %
    urlcolor=blue, %
    pdftoolbar=true, %
    pdfmenubar=true, %
    pdffitwindow=false, %
    pdfstartview={FitH} %
}
\begin{document}
%
%
%
%
%
\title{Reinforcement-Learning-Designed Field-Free Sub-Nanosecond Spin-Orbit-Torque Switching}
\date{\today}
\author{Yuta Igarashi}
\affiliation{Department of Electrical Engineering, Electronics, and Applied Physics, Saitama University, Saitama, 338-8570, Japan}

\author{Junji Fujimoto}
\affiliation{Department of Electrical Engineering, Electronics, and Applied Physics, Saitama University, Saitama, 338-8570, Japan}
\email{jfujimoto@mail.saitama-u.ac.jp}

\begin{abstract}
We demonstrate deterministic, field-free magnetization reversal of a single-domain nanomagnet within 300~ps under a current density of $3 \times 10^{10}~\mathrm{A/m^2}$ by coupling reinforcement learning~(RL) to the Landau-Lifshitz-Gilbert equation with the spin-orbit torques~(SOTs).
The RL agent autonomously discovers a current waveform that minimizes the magnetization trajectory path and exploits a precessional shortcut enabled by the field-like SOT and hard-axis anisotropy.
From the learned pulse, we extract a clear physical picture of the dynamics and develop a model-based analytical framework that establishes a lower bound on the switching time.
The control strategy remains robust across a wide range of damping constants and is stabilized against thermal fluctuations at higher current densities.
We also discuss feasible experimental implementations for the precessional switching.
\end{abstract}
\maketitle

Nanomagnets controlled by electric currents promise non‑volatile memory cells and spin-torque oscillators that switch orders of magnitude faster than charge‑based CMOS devices, provided their magnetization reversal is deterministic and energy-efficient.
Yet the ultimate speed limit for such field‑free reversal is still unknown, which restricts next‑generation memories, neuromorphic computing, and radio‑frequency spin‑logic.

A promising route is ballistic reversal, or previously called the precessional switching~\cite{back1998,back1999}, in which a single impulse drives the magnetic moment through roughly 180$^{\circ}$ of precession and lets it settle in the opposite direction.
Pioneering experiments realized this motion with carefully shaped magnetic-field pulses~\cite{schumacher2003,schumacher2003a}, and the concept was later extended to spin-orbit torques~(SOTs)~\footnote{Strictly speaking, these torques should be designated ``spin Hall torques'' to distinguish them from the original spin-orbit torques introduced in Refs.~\cite{manchon2008,manchon2009}. The torques customarily labeled ``SOTs'' actually stem from the combined action of the spin Hall effect~\cite{sinova2015} and spin-transfer torque~\cite{slonczewski1996,berger1996}. Nevertheless, in keeping with prevailing convention, we refer to them as ``spin-orbit torques'' throughout this paper.}, where a heavy-metal underlayer converts an electric current into field-like and damping-like SOTs~\cite{lee2018,polley2023}.
However, each demonstration has relied on manually tuned waveforms, making it challenging to utilize ballistic reversal to other materials, damping constants, or device geometries.

Reinforcement learning~(RL), well known for mastering Go~\cite{silver2016} and autonomous driving~\cite{kiran2022}, offers a way to automate such fine-tuning.
An RL agent improves its strategy through trial and feedback alone, without the need for labelled data or a perfect model.
Recently, RL has demonstrated record-high fidelities in quantum gate design~\cite{metz2023} and enabled efficient searches in many-body physics~\cite{yao2021,lu2023}.
In spintronics, RL has so far been used to reduce failure rates in perpendicularly magnetized cells~\cite{ender2022}; its capacity for ultrafast (less than hundreds of picoseconds) reversal remains unexplored.

\begin{figure}
\centering
\includegraphics[width=\linewidth]{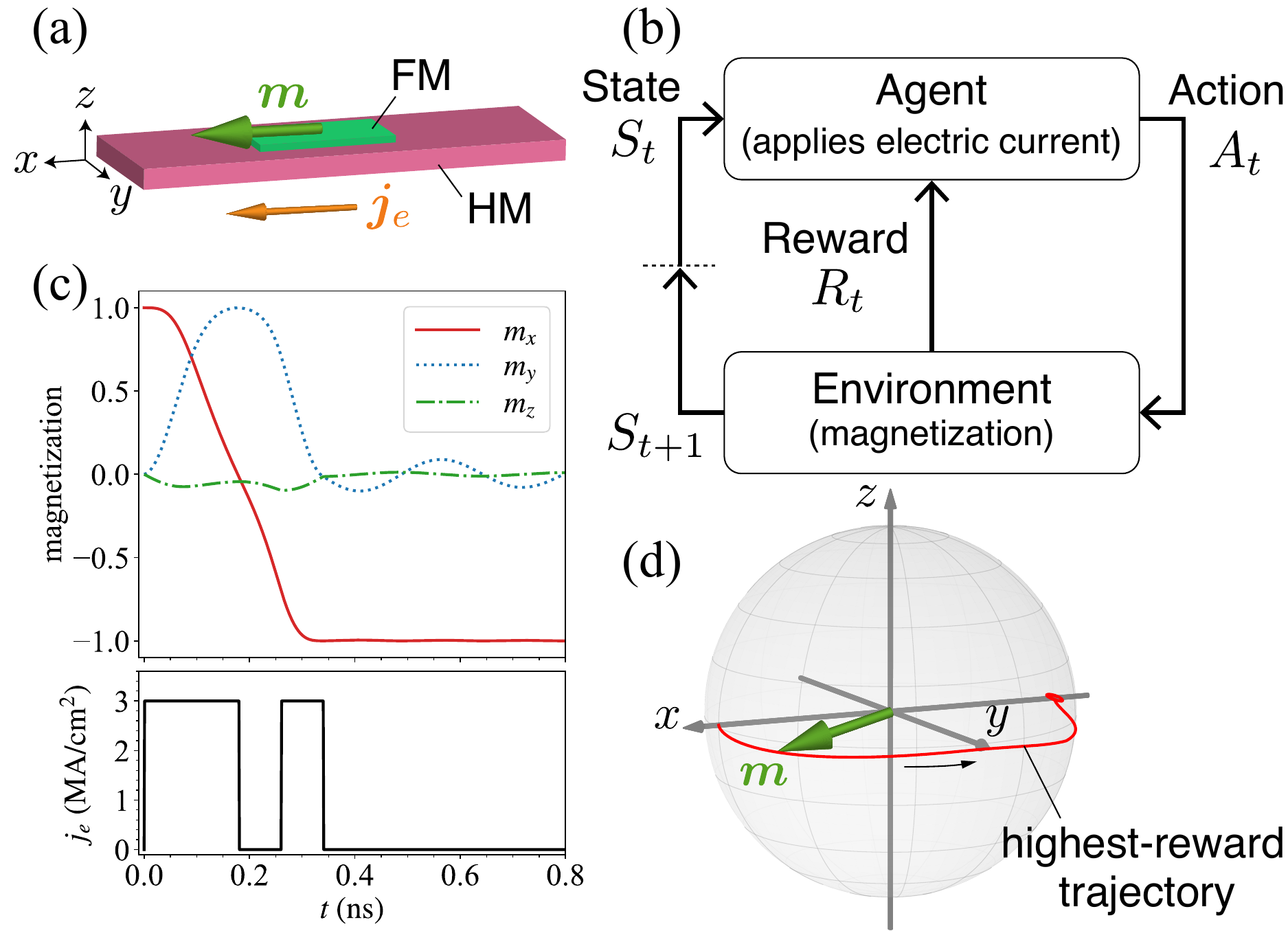}
\caption{\label{fig:1}
(a)~Schematic of a bilayer system composed of a ferromagnetic metal~(FM) and a heavy metal~(HM).
We consider the so-called type-$x$ configuration~\cite{fukami2016}, where the magnetic easy axis lies in-plane, aligned with the current direction.
(b)~Reinforcement learning~(RL) framework for magnetization reversal.
At each time step $t$, the magnetization state $S_t$ is observed, and the agent selects an action $A_t$ (whether to apply current).
The system evolves under this action to a new state $S_{t+1}$, based on which the agent receives a reward $R_t$.
(c)~Magnetization dynamics (top) driven by the highest-reward current waveform (bottom) at $T = 0$.
Reversal is achieved within 300~ps under a current density of $3~\mathrm{MA/cm^2}$.
(d)~Magnetization trajectory corresponding to the highest-reward reversal found by the RL agent.
}
\end{figure}

Here, we embed a deep Q‑network~(DQN) agent~\cite{mnih2013} into numerical simulations governed by the Landau-Lifshitz-Gilbert equation for single-domain nanomagnets modeling W/CoFeB bilayer systems~(Fig.~\ref{fig:1}~(a)).
We show that the DQN agent learns how to reverse the magnetization~(Fig.~\ref{fig:1}~(b)) and finds a time‑dependent waveform that switches the magnetization within $300~\mathrm{ps}$ using a current density of $3~\mathrm{MA/cm^2}$~(Fig.~\ref{fig:1}~(c) and (d)).
We identify the key ingredients of this field-free, ballistic process, the interplay between the field-like SOT and the hard-axis anisotropy, and clarify their respective roles.
Based on the physical picture, we further estimate the lower bound time for the ballistic reversal.
Our findings provide a universal control route toward ultrafast and energy-efficient memories, neuromorphic computing, radio-frequency spin logic, and other emerging spintronic platforms.

Now we introduce the theoretical framework of embedding the RL into magnetization dynamics.
We consider a macrospin model for the so-called type-$x$ configuration~\cite{fukami2016}, where the magnetic easy axis lies in-plane, aligned with the current direction, and the Landau-Lifshitz-Gilbert~(LLG) equation with the SOTs~\cite{liu2011,taniguchi2015},
\begin{align}
\dot{\bm{m}}
    & = - \gamma \bm{m} \times \bm{H}_{\mathrm{eff}}
        + \alpha_{\mathrm{G}} \bm{m} \times \dot{\bm{m}}
\notag \\ & \hspace{1em}
        - \gamma H_s \bm{m} \times (\hat{y} \times \bm{m})
        - \gamma \beta H_s \bm{m} \times \hat{y}
\label{eq:LLG}
,\end{align}
where $\bm{m} = (m_x, m_y, m_z)$ is the unit vector of magnetization, $\gamma$ is the gyromagnetic ratio, $\alpha_{\mathrm{G}}$ is the Gilbert damping constant.
The two terms in the second line of Eq.~(\ref{eq:LLG}) describe the damping-like and field-like SOTs, respectively, where $H_s$ is the torque strength given by
\begin{align}
H_s
    & = \frac{\hbar}{2 e} \frac{\theta_{\mathrm{SH}}}{M_{\mathrm{s}} l_z} j_e
\end{align}
with $\theta_{\mathrm{SH}}$ being the spin Hall angle of HM, $M_{\mathrm{s}}$ the saturation magnetization, $l_z$ the thickness of FM, and $j_e$ the current density, which is the changeable value by the RL agent. 
The parameter $\beta$ represents the ratio of the field-like torque to the damping-like torque.
We set $\theta_{\mathrm{SH}} = -0.25$ and $\beta = -3$ from experiments on W/CoFeB systems~\cite{pai2012,park2023}, throughout this paper.
The effective magnetic field $\bm{H}_{\mathrm{eff}}$ consists of the demagnetization field $\bm{H}_{\mathrm{demag}}$, the oersted field $\bm{H}_{\mathrm{Oe}}$ due to the electric current, and the thermal noise field $\bm{H}_{\mathrm{th}}$.
We first perform the simulations at zero temperature, $T = 0$, and later discuss the effects of the thermal noise field.
The demagnetization field is given by $\bm{H}_{\mathrm{demag}} = - \mu_0 M_{\mathrm{s}} (N_x m_x \hat{x} + N_y m_y \hat{y} + N_z m_z \hat{z})$ with the vacuum permeability $\mu_0$, where the demagnetization factor $(N_x, N_y, N_z) \simeq (0.0154, 0.0316, 0.952)$ is determined from the FM rectangle shape with $l_x = 100~\mathrm{nm}$, $l_y = 50~\mathrm{nm}$, and $l_z = 1~\mathrm{nm}$, based on the formula by Aharoni~\cite{aharoni1998} (for the exact value of the demagnetization factor, see the companion paper~\cite{companion}).
We emphasize that $\bm{H}_{\mathrm{demag}}$ shows the strong hard-axis anisotropy in the $z$ direction.
The oersted field is determined by Amp\`{e}re's law; $\bm{H}_{\mathrm{Oe}} = (\mu_0 j_e d_{\mathrm{HM}}/2) \hat{y}$, where we set $d_{\mathrm{HM}} = 5~\mathrm{nm}$ in this paper.
Note that we do not apply any external magnetic field.

We reformulate the LLG dynamics with the SOTs, the underlying model governing the time evolution of magnetization, as a control problem of a Markov decision process to apply the RL scheme.
In this framework, we adopt the DQN algorithm.
The magnetization vector is treated as the state of the environment, while the external electric current, which can be either applied or not applied, represents the agent’s discrete actions.
The reward function is designed as $R_t = - \{ m_x (t) \}^3$, to promote rapid and reliable switching.
This cubic form ensures that the reward becomes significantly higher as $m_x$ approaches $-1$, encouraging trajectories that rapidly drive the magnetization along the negative $x$-direction.

\begin{figure}
\centering
\includegraphics[width=\linewidth]{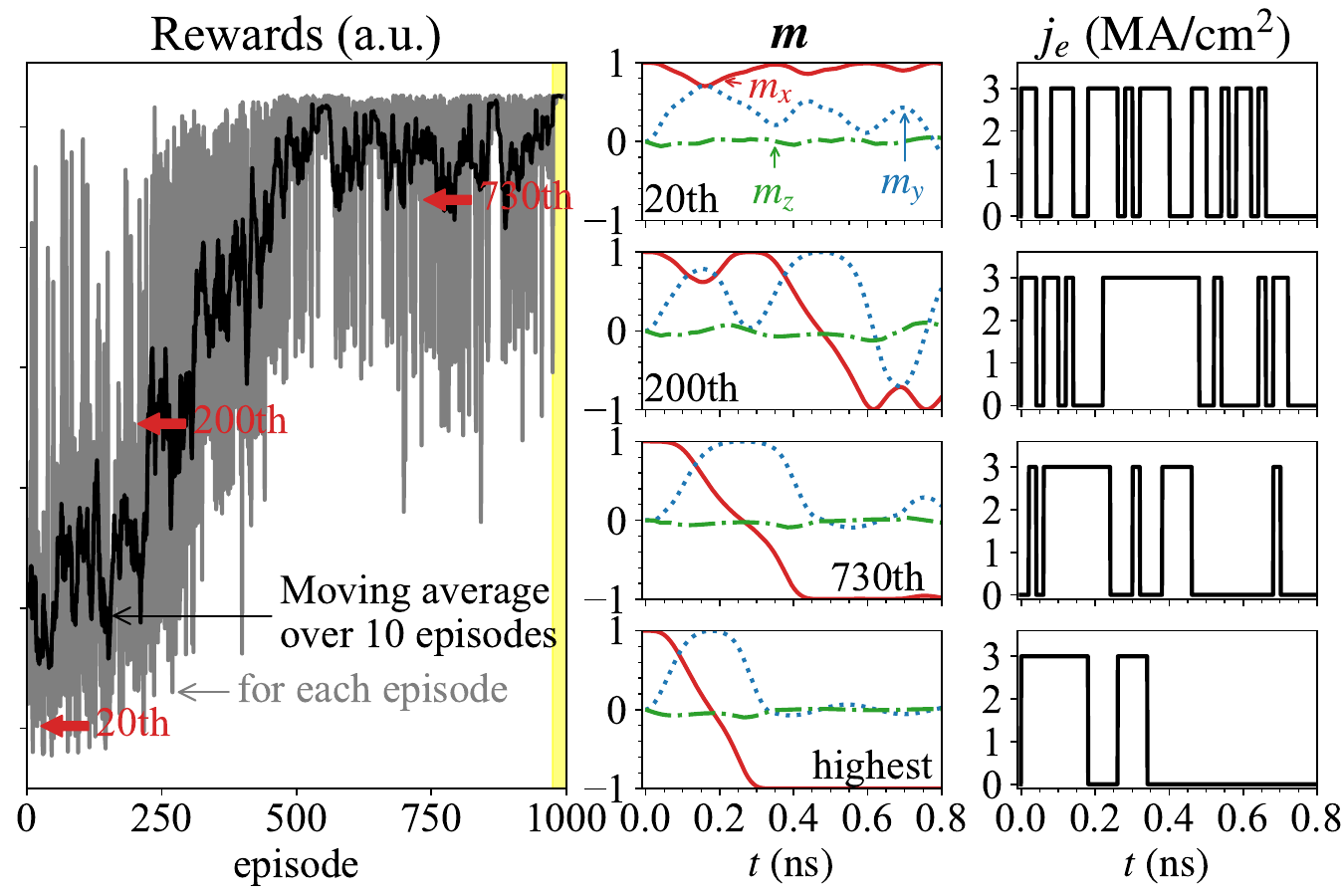}
\caption{\label{fig:2}
(Left)~Episode dependence of the total rewards.
The overall learning curve exhibits a general trend of improvement, although the reward in each episode fluctuates due to the exploratory nature of the DQN algorithm.
In the last 25~episodes (marked by the yellow rectangle), the exploration rate was set to zero, and hence the agent performed only exploitation; as a result, the rewards became stable.
(Center)~Magnetization dynamics for the 20th, 200th, and 730th episodes and the episode yielding the highest total reward.
(Right)~The corresponding current waveforms.
The period of a single episode is set to 0.8~ns, and the action interval is set to 20~ps.
The simulation is performed for $\alpha_{\mathrm{G}} = 0.01$ and $M_{\mathrm{s}} = 750~\mathrm{kA/m}$ (i.e., $\mu_0 M_{\mathrm{s}} = 0.94~\mathrm{T}$) at zero temperature $T = 0$.
}
\end{figure}
\begin{figure}
\centering
\includegraphics[width=\linewidth]{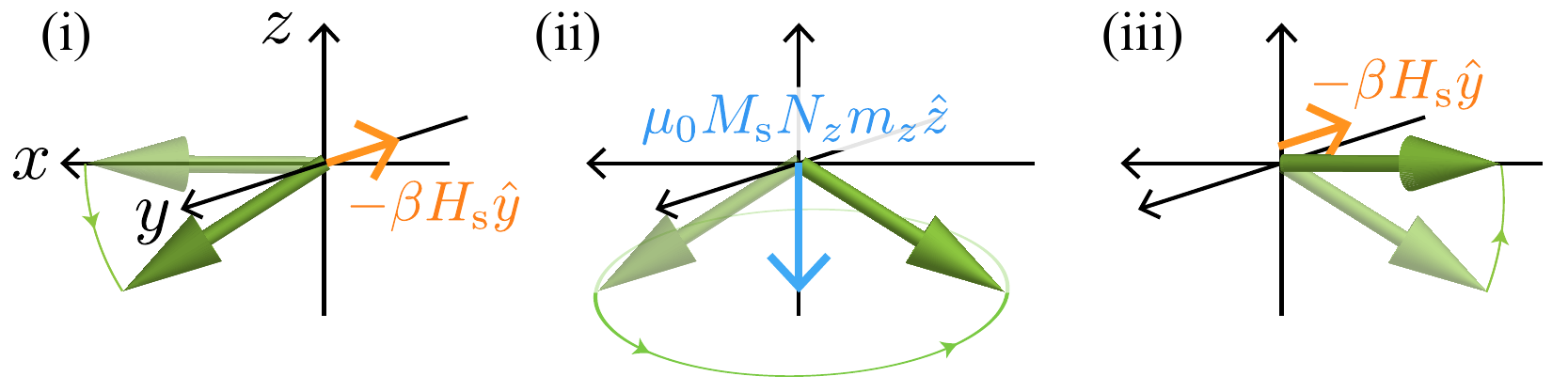}
\caption{\label{fig:3}Schematic of ballistic magnetization reversal driven by the field-like SOT in the presence of a strong hard-axis anisotropy.
(i)~The magnetization, initially along $+x$, is tilted out of plane by the field-like SOT, yielding $m_z<0$.
(ii)~The sizeable anisotropy field then launches a rapid precession of the magnetization around the $z$ axis.
(iii)~When the precession carries the magnetization to the $-x$ direction, a second current pulse is injected. The field-like SOT reduces the out-of-plane component ($m_z \simeq 0$), arrests the precession, and stabilizes the reversed state.}
\end{figure}
Figure~\ref{fig:2} provides an overview of the training process, including the episode-dependent total rewards and representative examples of magnetization dynamics and current profiles.
The overall learning curve shows a general trend of improvement, although the cumulative reward exhibits fluctuations due to the exploratory nature of the DQN algorithm.
In the early phase of training, the reward fluctuates significantly due to random exploration, but a gradual increase in performance can be observed as the agent accumulates experience.
After approximately 500~episodes, the reward stabilizes within a high-performance regime.
In the final 25 episodes, the exploration rate was set to zero so that the agent exploited the learned policy exclusively, during which the total reward consistently reached its highest values, indicating stable and optimal behavior.
The representative current profiles and magnetization trajectories shown in Fig.~\ref{fig:2} illustrate the qualitative shift in policy behavior, from erratic and inefficient pulses in early episodes to smooth and effective reversal in the later stages.
We further comment on the stability of the learning process, noting that convergence was robust across multiple runs and that the overall performance was not highly sensitive to moderate variations in hyperparameters.

Figure~\ref{fig:3} gives a physical picture of the discovered ballistic reversal, which is a different reversal mechanism from the field-induced mechanism~\cite{schumacher2003,schumacher2003a}, the heat-assisted mechanism~\cite{wernsdorfer1996,koch2000}, and the microwave-assisted mechanism~\cite{thirion2003}.
The reversal process consists of three stages as illustrated in Fig.~\ref{fig:3}: the tilting stage, the half-precession stage, and the termination stage.
(i)~Tilting stage: an applied current pulse generates a field-like SOT that tilts the magnetization from the $+x$ axis toward the $-z$ direction.
The process is essentially dissipationless because the pulse duration is much shorter than the damping time scale.
(ii)~Half-precession stage: the hard-axis anisotropy supplies an effective field $\bm{H}_{\mathrm{demag}} \sim - \mu_0 M_{\mathrm{s}} N_z m_z \hat{z}$ that exceeds a sub-tesla, launching a nearly rigid precession of the magnetization about the $z$ axis.
The half-precession period, set by $\pi / |\gamma H_{\mathrm{demag}}|$, lies well below 300~ps under the present parameters.
(iii)~Termination stage: when the precession angle reaches $180^{\circ}$ (magnetization along $-x$), a second, precisely-tuned current pulse is injected.
Its field-like SOT forces $m_z \simeq 0$, stopping the precession and locking the magnetization in the reversed state.
Since the reversal trajectory follows the constant-energy orbit dictated by the anisotropy field, energy dissipation is minimized; deterministic switching is therefore achieved within a single half-precession, with no need for thermal activation or external assist fields.
In the above description, magnetic damping is not essential, which implies that the reversal mechanism is robust against a wide range of Gilbert damping parameters.

We now estimate the lower bound of the reversal time for ballistic magnetization reversal by applied current $j_e$.
For each of the three stages introduced above, we construct simple macrospin models and analyze them (for detailed derivations, see the companion paper~\cite{companion}).
In the tilting stage, the characteristic time scale is denoted by
\begin{align*}
t_1
    & \simeq \frac{\sqrt{2}}{\omega_2} \sqrt{
        - 1 + \sqrt{
            1
            + \frac{\omega_2^2}{(\gamma \beta H_{\mathrm{s}})^2} \left( \frac{N_y - N_x}{N_z - N_x} \right)
                \left( 1 + \frac{\delta \epsilon}{N_y - N_x} \right)
        }
    }
,\end{align*}
whereas the half-precession stage lasts roughly $t_2 \simeq 2 K (k) / u \omega_2$, where $\delta \epsilon \simeq \alpha_{\mathrm{G}} \sqrt{(N_z - N_y) (N_y - N_x)}$, $K (k)$ is the complete elliptic integral of the first kind, $k = 1 - (l/u)^2$ with $l$ and $u$ is the lower and upper bounds of $m_z$ in the half precession stage, and $\omega_2 = 
\mu_0 M_{\mathrm{s}} \sqrt{(N_z - N_y) (N_z - N_x)}$ is the precession frequency.
Finally, in the termination stage, the dominant time scale is again $t_1$, mirroring that of the tilting stage.
Hence, the lower bound of the reversal is expressed by $T = 2 t_1 + t_2$.

\begin{figure}
\centering
\includegraphics[width=\linewidth]{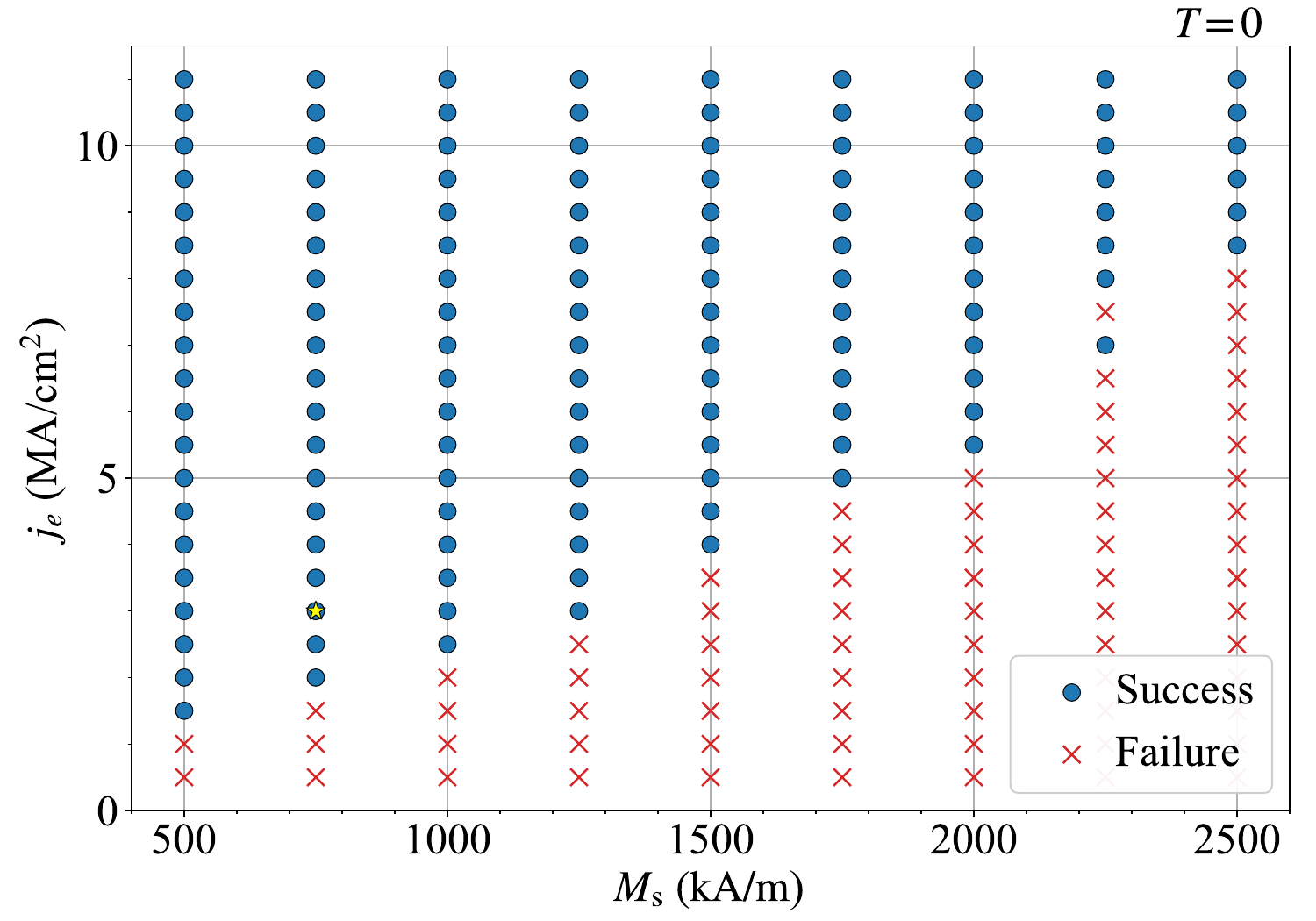}
\caption{\label{fig:4}
Success/failure map of the ballistic magnetization reversal as a function of the saturation magnetization $M_{\mathrm{s}}$ and the applied current density $j_e$ at zero temperature $T = 0$.
The threshold current density rises monotonically with increasing $M_{\mathrm{s}}$.
The yellow star marks the parameter set in Figs.~\ref{fig:1} and \ref{fig:2}.
The definitions of success and failure are described in the main text.
The success for $M_{\mathrm{s}} = 2250~\mathrm{kA/m}$ with $j_e = 7.0~\mathrm{MA/cm^2}$ is by the swinging mechanism (see the companion paper~\cite{companion} for the detail).
Stable deterministic ballistic reversal at room temperature requires a higher current density (see Fig. \ref{fig:5}).
}
\end{figure}
Figure~\ref{fig:4} presents the zero-temperature switching phase diagram in the ($M_{\mathrm{s}}$, $j_e$) plane.
For each point, the DQN agent learns through the 1000-episode trials as shown in Fig.~\ref{fig:2}, where the single episode period is set to 0.8~ns; blue circles denote successful ballistic reversal, whereas red crosses indicate failure.
We define success and failure as follows.
During $0 \le t \le 0.8~\mathrm{ns}$, the current is modulated according to the learned optimal policy, after which ($0.8 < t \le 2~\mathrm{ns}$) the LLG equation is integrated with no applied current.
A trajectory is classified as `success' if the final $x$-component of the magnetization satisfies $m_x < -0.85$ during the last 0.5~ns; otherwise, it is labeled `failure'.
The threshold current density $j_{\mathrm{th}}$, defined by the change from failure to success, rises almost linearly with the saturation magnetization over the investigated range from $500$ to $2500~\mathrm{kA/m}$.
This monotonic trend originates roughly from the dependence of the torque strength; $H_{\mathrm{s}} \propto j_e/M_{\mathrm{s}}$.
For the material parameters used in Figs.~\ref{fig:1} and \ref{fig:2}, $M_{\mathrm{s}} = 750~\mathrm{kA/m}$; yellow star in Fig.~\ref{fig:4}, the full simulation yields $j_{\mathrm{th}} \simeq 2~\mathrm{MA/cm^2}$.
At room temperature, the stable value to perform the ballistic reversal is approximately tripled (see below), indicating that reliable ballistic operation in practical devices will require larger drive currents than the ideal values shown here.

Near the threshold current density, the RL agent finds waveforms that reverse the magnetization by swinging it.
Indeed, the success for $M_{\mathrm{s}} = 2250~\mathrm{kA/m}$ with $j_e = 7.0~\mathrm{MA/cm^2}$ is by the swinging mechanism.
A single pulse cannot induce magnetization reversal; therefore, the agent applies multiple current pulses in a precisely timed sequence.
These pulse sequences swing the magnetization larger, resulting in the magnetization reversal.
For additional information on the swinging mechanism, see the companion paper~\cite{companion}.

Here, we discuss two effects of thermal fluctuations on ballistic magnetization reversal.
The thermal noise field is treated as $\bm{H}_{\mathrm{th}} = h_{\mathrm{th}} \bm{\eta} (t)$ with the time-dependent random vector $\bm{\eta} (t)$ from the standard normal distribution and the field strength
\begin{align}
h_{\mathrm{th}}
    & = \sqrt{\frac{2 \alpha_{\mathrm{G}} k_{\mathrm{B}} T}{\gamma M_{\mathrm{s}} V \varDelta t}}
,\end{align}
where $V = l_x l_y l_z$ is the volume of FM, and $\varDelta t$ is the time step of LLG time evolution.

First, thermal noise affects the RL process: the stochastic field introduces randomness into the magnetization dynamics and can prevent the agent from discovering the optimal policy. The magnitude of this effect can be estimated by comparing the field-like SOT strength with the thermal noise strength.
Assuming $T = 300~\mathrm{K}$, $\alpha_{\mathrm{G}} = 0.01$, $M_{\mathrm{s}} = 750~\mathrm{kA/m}$, $V = 5 \times 10^{-24}\mathrm{m^{3}}$, and time step $\Delta t = 1\mathrm{ps}$, the noise strength is $h_{\mathrm{th}} \simeq 10~\mathrm{mT}$.
The field-like SOT strength is $\beta H_{\mathrm{s}}/j_e \simeq 3.2~\mathrm{mT}$. Hence, at $j_{e} \approx 3~\mathrm{MA/cm^2}$, the thermal field is comparable to the field-like SOT strength, and learning fails to converge: after 1000 episodes, the reward remains unstable even when the exploration rate is set to zero during the last 25 episodes.
For additional results, see the companion paper~\cite{companion}.
By contrast, when $j_{e} \approx 5~\mathrm{MA/cm^2}$, the SOT strength exceeds the thermal noise, the learning process succeeds, and the total rewards in the final 25 episodes are stable.

\begin{figure}
\centering
\includegraphics[width=\linewidth]{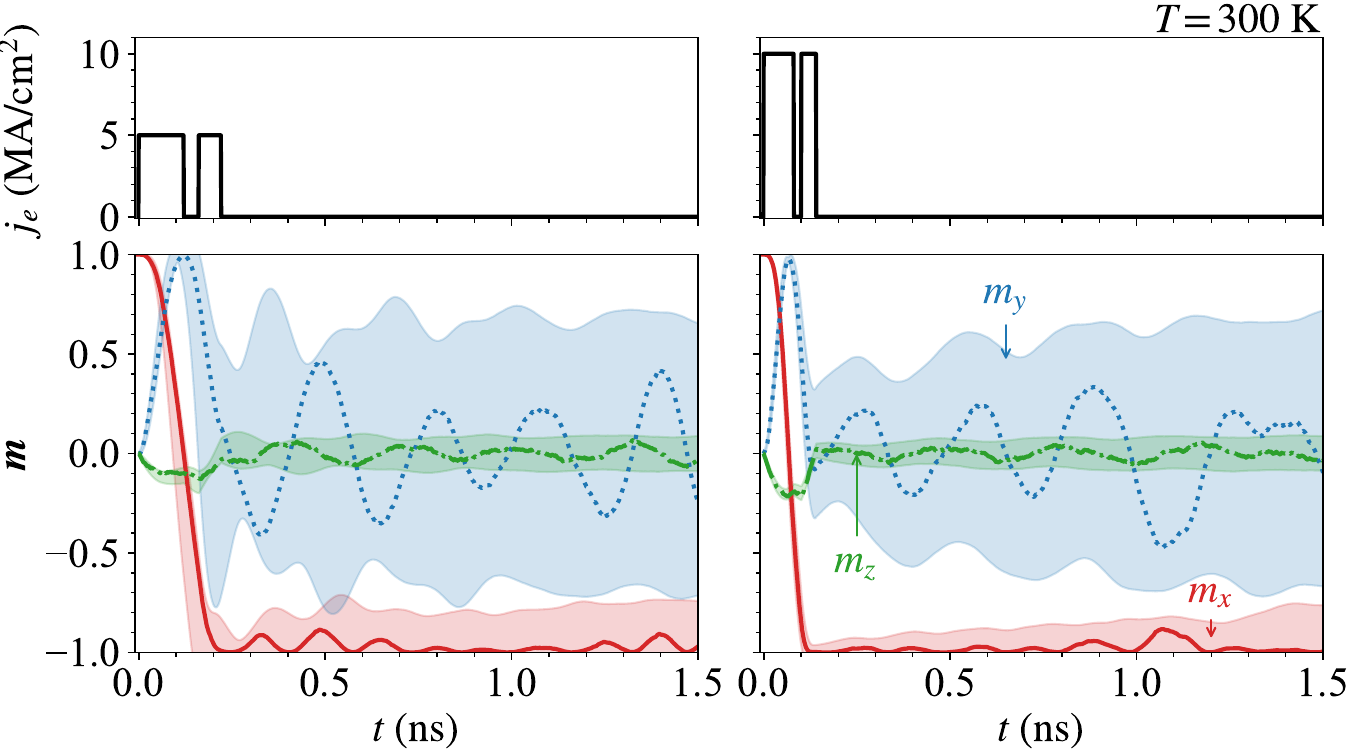}
\caption{\label{fig:5}
(Upper) Optimized current pulse waveforms; (lower) the corresponding average trajectories of the magnetization dynamics at $T = 300~\mathrm{K}$, accompanied by the 99\% confidence interval (assuming a normal distribution), representing statistical variation.
The confidence interval is calculated from 1000 simulations of the LLG equation for each current pulse waveform.
A narrower confidence interval at higher current density ($j_e = 10~\mathrm{MA/cm^2}$) indicates that the ballistic magnetization reversal is more stable.
}
\end{figure}
Second, even the highest-reward current waveform found by the RL agent can make the ballistic reversal unstable, because thermal noise generates a random torque.
In other words, applying the same waveform does not guarantee identical magnetization dynamics in different trials.

Figure \ref{fig:5} summarizes the stability of the highest-reward waveforms for current densities $j_e = 5~\mathrm{MA/cm^2}$ and $j_e = 10~\mathrm{MA/cm^2}$.
The upper panels display the optimized current-pulse waveforms found by the RL agent; the lower panels show the corresponding trajectories of the magnetization dynamics together with the 99\% confidence interval (assuming a normal distribution), which quantifies the statistical variation.
Each confidence interval is obtained from 1000 stochastic LLG simulations for the given waveform.
The markedly narrower confidence band at the higher current density ($j_e = 10~\mathrm{MA/cm^2}$) demonstrates that the ballistic reversal is more stable under stronger SOT fields.
Note, however, that the reversal becomes unstable again at still higher currents, e.g., $j_e \sim 20 \approx \mathrm{MA/cm^2}$, because the second pulse grows too short and its timing becomes critical for stopping the precessional motion.

Finally, we discuss a potential experimental implementation for ballistic magnetization reversal.
As depicted in Fig.~\ref{fig:3}, three ingredients are crucial: (i) a sizable field-like SOT, (ii) a large out-of-plane hard-axis anisotropy, and (iii) precise timing of the electric current pulse that terminates the precessional motion.
The optimum pulse width and duration can be estimated directly from the present theory.
A stronger field-like SOT tilts the magnetization farther toward the out-of-plane direction, enhancing the anisotropy field and accelerating the precession.
A promising materials platform is a W/CoFeB/MgO system, where the large spin-orbit coupling of W generates a pronounced field-like SOT~\cite{park2023}.
The type-$x$ geometry proposed in Ref.~\cite{fukami2016} is suitable for realizing the ballistic reversal.
Time-resolved magneto-optic Kerr microscopy with sub-picosecond resolution can trace the coherent precession and verify the predicted half-period switching.
We emphasize that, unlike previous quasi-static switching experiments, the present protocol does not require a bias field; the single-parameter control over pulse area can achieve the energy-efficient reversal.
Overall, the proposed setup is compatible with state-of-the-art nanofabrication and ultrafast electronics and thus provides a realistic route to test the ballistic magnetization reversal scenario advanced in this work.

\begin{acknowledgments}
The authors thank Y.~Nozaki and K.~Kutsuzawa for valuable comments.
This work was supported by JSPS KAKENHI Grant Numbers JP22K13997 and JP25K07218, by JST-CREST Grant Number JPMJCR19J4, and by JST-Mirai Program Grant Number JPMJMI20A1.
\end{acknowledgments}

\bibliography{reference}

\begin{thebibliography}{28}
\bibitem [{com()}]{companion}%
  \BibitemOpen
  \href@noop {} {\bibinfo  {journal} {[URL will be inserted by publisher] for detailed derivations and additional results, see the companion paper.}\ }\BibitemShut {NoStop}%
\end{thebibliography}
\end{document}